\documentclass[aps,twocolumn,showpacs,prl,amsmath,amssymb]{revtex4}
\usepackage{times}
\usepackage{graphicx}
\usepackage{dcolumn}
\usepackage{bm}

\begin{document}
\title{Giant non-adiabatic effects in layer metals: Raman spectra of intercalated
graphite explained}
\author{A. Marco Saitta}
\author{Michele Lazzeri}
\author{Matteo Calandra}
\author{Francesco Mauri}
\affiliation{ IMPMC, Universit\'es Paris 6 et 7, CNRS, IPGP, 140
rue de Lourmel, 75015 Paris, France
}
\date{\today}
\begin{abstract}
The occurrence of non-adiabatic effects in the vibrational
properties of metals have been predicted since the 60's, but
hardly confirmed experimentally. We report the first fully
\emph{ab initio} calculations of non-adiabatic frequencies of a
number of layer and conventional metals. We suggest that
non-adiabatic effects can be a feature of the vibrational Raman
spectra of any bulk metal, and show that they are spectacularly
large (up to 30\% of the phonon frequencies) in the case of layer
metals, such as superconducting $MgB_2$, $CaC_6$ and other
graphite intercalated compounds. We develop a framework capable to
estimate the electron momentum-relaxation time of a given system,
and thus its degree of non-adiabaticity, in terms of the
experimentally observed frequencies and linewidths.
\end{abstract}
 \maketitle


The adiabatic Born-Oppenheimer approximation (ABOA) is the state
of the art in first-principles calculations of vibrational
properties in solids. In metals, even if in principle unjustified,
this approximation generally leads to phonon dispersions in very
good agreement with experimental data \cite{BaroniRMP}. Indeed
violations of the ABOA are hardly visible in solids. Engelsberg
and Schrieffer suggested \cite{Engelsberg63} that non-adiabatic
(NA) effects could lead to a significant renormalization of
zone-center optical phonon frequencies. An intense effort have
been devoted to detect measurable NA effects in metals, mainly by
Raman spectroscopy \cite{Maksimov95,Ponosov98,Bolotin01}, however
the detected NA phonon-frequencies-renormalizations are typically
smaller than some percent of the adiabatic phonon frequency.

Recently it has been shown that NA effects are crucial to
interpret the dependence of Raman spectra on doping
\cite{LazzeriMauriPRL06,Pisana07} in graphene and on doping
\cite{Caudal07,Das07} and on diameter and
temperature\cite{Piscanec07} in nanotubes. However, despite being
a central issue for the physics of graphene-based systems, the NA
phonon frequency shift $\Delta \omega$ (the difference between
adiabatic and NA phonon frequencies), although  measurable, is
less than 1\% of the adiabatic phonon-frequency. Moreover, these
systems have lower dimensionality and a peculiar electronic
structure, it is then unclear whether sizeable NA effects can
actually be observed in truly three-dimensional bulk metals.

Layer metallic materials, such as Graphite Intercalated Compounds
(GICs), are three dimensional (3D) metals possessing a
considerable anisotropy along one direction. Variation of the
interlayer distance by intercalation or applied pressure allows to
bridge the gap between two-dimensional monolayer systems (such as
graphene) and truly three-dimensional systems
\cite{CalandraPRB06}. Thus these metals are ideal to judge the
role of reduced dimensionality in determining non-adiabatic
effects. CaC$_6$ is also an 11.5 K
superconductor\cite{WellerNatPhys05} with an intermediate
electron-phonon coupling $\lambda\approx 0.83$
\cite{CalandraPRL05}. Most interestingly, the recent measurement
of CaC$_6$ Raman spectrum~\cite{Nancy07} shows that the phonon
frequencies related to in-plane C vibrations are almost
80$~cm^{-1}$ larger than those obtained from density-functional
theory (DFT) adiabatic calculations. This result is puzzling since
in graphite the adiabatic result for the E$_{2g}$ phonon frequency
is 1577$~{\rm cm}^{-1}$, in excellent agreement with the
experimental value of 1582$~{\rm cm}^{-1}$.

In this Letter we develop a first-principles theoretical framework
to calculate the magnitude of NA effects on zone-center
optical-phonons in metals. We identify the general conditions for
having sizeable adiabatic effects in 3D bulk metals and analyze
the experimental constraints that can hinder the observation of NA
effects. We demonstrate that {\it the occurrence of NA effects is
not limited to reduced dimensionality but it is a general
properties of metals}. We apply our approach to GICs,  MgB$_2$ and
a few bulk metals, finding giant NA effects. To the best of our
knowledge, our work is the first systematic implementation and
study of non-adiabatic effects in a first principles framework.

Non adiabatic effects due to the treatment of the electron-phonon
coupling in the Migdal approximation (neglecting of vertex
corrections in the adiabatic limit) are usually of the order
$\sqrt{ m/M}$, where $m$ and $M$ are the electronic and ionic mass
respectively, and thus generally very small. Engelsberg and
Schrieffer ~\cite{Engelsberg63} showed that NA renormalization of
adiabatic phonon frequencies, {\it a larger effect, unrelated to
the neglecting of vertex corrections, and not of the order of
$\sqrt{ m/M}$}, can be observed if the following two conditions
are satisfied:
\begin{eqnarray}
|\textbf{q}\cdot {\bf v_F}|&\ll& \omega  \label{eq1}\\
\hbar\omega&\gg&\sigma \label{eq2}
\end{eqnarray}
where ${\bf v_F}$ is the Fermi velocity, \textbf{q} is the phonon
wavevector, $\omega$ is the phonon frequency, $\sigma=\hbar/\tau$
and $\tau$ is the electron momentum-relaxation time (Drude) of the
electrons near the Fermi surface due to all possible
momentum-exchange scattering mechanisms.

The first condition is verified for optical phonons of small
wavevector whose phase velocity is larger than the electronic
Fermi velocity. Even if in principle Eq. (\ref{eq1}) can be
fulfilled in any system, provided {\bf q} is small enough, in
practice the penetration length of the laser light limits the
smallest exchanged {\bf q} in Raman experiments. The fact that Eq.
(\ref{eq1}) is experimentally difficult to satisfy in 3D metals
explains why the observation of NA effects in these systems has
been somehow disappointing.

The second condition states that the electron momentum-relaxation
time must be much larger than the phonon period. This implies that
the electron band-population dynamic is too slow to follow the
atomic motion and thus the dynamic is non-adiabatic
\cite{Pisana07}. Eq. (\ref{eq2}) is usually verified in pure and
well crystallized samples at low temperature. Thus Eq. (\ref{eq1})
is the main limitation to the observation of NA effects.

This limitation can be circumvented considering system having
small \textbf{v$_F$} along certain directions. For example layer
metals are usually characterized by a small ${\bf v_F}$ component
perpendicular to the layers. Since the samples are usually cleaved
parallel to the layers and Raman experiments are performed
perpendicularly to the freshly cleaved surface,  the scalar
product \textbf{q}\textbf{$\cdot$}\textbf{v$_F$} is small. It is
instead also evident that condition (\ref{eq1}) is easily verified
in 2D or 1D systems, since Raman experiments are performed with an
incident light of wavevector \textbf{q} perpendicular to the
sample spatial dimension(s) and, thus, to \textbf{v$_F$}. In fact,
NA effects have been consistently observed in doped graphene and
nanotubes in the last few years.

NA effects can be taken into account by applying time-dependent
perturbation theory to DFT. {\it Neglecting the electron-momentum
scattering rate ($\sigma=0$)}, the dynamical matrix (${\cal D}$)
for a phonon at {\bf q=0} (${\bf \Gamma}$) and the first-order
variation of the electronic charge density ($\Delta n$) are
\cite{footnote1}:
\begin{eqnarray}
{\cal D}_{\bf \Gamma}(\omega)&=&\frac{2}{N_{\bf k}}\sum_{{\bf
k}n,m\ne n} \frac{|D_{{\bf k}m,{\bf k}n}|^2 [f_{{\bf k}m}-f_{{\bf
k}n}]} {\epsilon_{{\bf k}m}-\epsilon_{{\bf
k}n}+\hbar\omega} \nonumber \\
&+&\int n(\textbf{r})
\Delta^2 V^b(\textbf{r})d\textbf{r}\nonumber \\
 &-&\int \Delta n_{\bf \Gamma}(\textbf{r},\omega)K(\textbf{r},{\bf r^\prime})
 \Delta n_{\bf \Gamma}(\textbf{r},\omega)d\textbf{r}d{\bf r^\prime}
\label{eq3} \\
\Delta n_{\bf \Gamma}(\textbf{r},\omega) &=& \frac{2}{N_{\bf k}}
\sum_{{\bf k}n,m\ne n} \frac{ \langle{\bf k}n
|\textbf{r}\rangle\langle \textbf{r}|{\bf k}m\rangle D_{{\bf
k}m,{\bf k}n} [f_{{\bf k}m}-f_{{\bf k}n}]} {\epsilon_{{\bf
k}m}-\epsilon_{{\bf
k}n}+\hbar\omega}\nonumber \\
\label{eq4}
\end{eqnarray}
where $n(\textbf{r})$ is the charge density, the sum is performed
on $N_{\bf k}$ k-points, $D_{{\bf k}m,{\bf k}n}= \langle {\bf
k}m|\Delta V^{sc}|{\bf k}n\rangle$ is the deformation potential
proportional to the electron-phonon matrix element, $\Delta
V^{sc}$ and $\Delta^2 V^b$ are, respectively, the first derivative
of the Kohn-Sham potential and the second derivative of the bare
(purely ionic) potential with respect to the phonon displacement.
The kernel $K({\bf r},{\bf r}')$ is the second functional
derivative of the Hartree exchange and correlation potential
respect to the densities at ${\bf r}$ and ${\bf r'}$. Finally
$f_{{\bf k}n}$ is the Fermi function for a Bloch state $|{\bf
k}n\rangle $ having band energy $\epsilon_{{\bf k} n}$. The NA
frequencies has to be computed self-consistenty from
$\omega^{NA}=\sqrt{{\cal D_{\bf \Gamma}}(\omega^{NA})/M}$. Since
in Eqs.~(\ref{eq3}) and~(\ref{eq4}) only interband transitions
contribute, if
$\hbar\omega^{NA}\ll|\epsilon_{\textbf{k}e}-\epsilon_{\textbf{k}o}|$,
\emph{i.e.} if the phonon energy is much smaller than the direct
gap between empty ($\epsilon_{\textbf{k}e}$) and occupied
($\epsilon_{\textbf{k}o}$) states, ${\cal D_{\bf
\Gamma}}(\omega^{NA})\simeq{\cal D_{\bf \Gamma}}(0)$. We verified
numerically that in all systems considered in the present study
this approximation applies and that the results obtained with
${\cal D_{\bf \Gamma}}(\omega^{NA})$ and ${\cal D_{\bf
\Gamma}}(0)$ are indistinguishable.

The adiabatic frequency $\omega^{A}$ is as usual calculated within
static perturbation theory, as in Ref.~\cite{BaroniRMP}. Note that
$\omega^{A}\ne\sqrt{{\cal D_{\bf \Gamma}}(0)/M}$ since in the
adiabatic case the intraband ($m$=$n$) term is
present~\cite{LazzeriMauriPRL06} and $\Delta\omega=\omega^{NA} -
\omega^A$ is:
\begin{equation}
\hbar\Delta\omega \simeq\frac{1}{N_{\bf k}}\sum_{{\bf k}n}
\frac{\hbar|D_{{\bf k}n,{\bf k}n}|^2}{M\omega^A}
\delta(\epsilon_F-\epsilon_{{\bf
k}n})=n(\epsilon_F)\overline{g^2(\epsilon_F)} , \label{eq5}
\end{equation}
where $n(\epsilon_F)$ is the density of states at the Fermi level,
$g^2(\epsilon_f)=\hbar |D_{{\bf k}n,{\bf k}n}|^2/2M\omega^A$ is
the square electron-phonon matrix element due to intraband
transitions, and $\overline{g^2(\epsilon_F)}$ is the average of
$g^2(\epsilon_f)$ over the Fermi surface.

\begin{figure}
\centerline{\includegraphics[width=8.25cm]{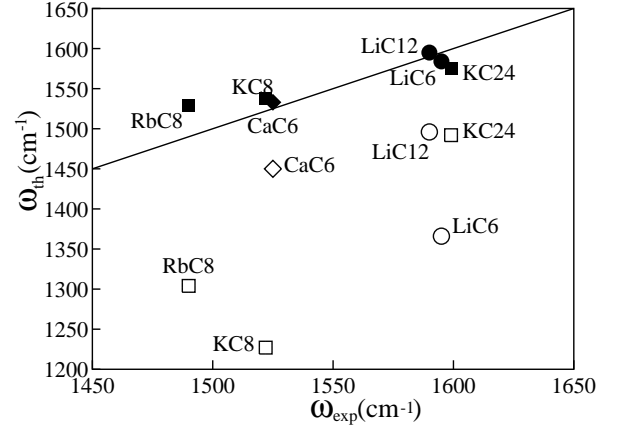}}
\caption{Calculated adiabatic (empty symbols) and non-adiabatic
(filled symbols) E$_{2g}$ phonon frequencies of several GICs,
compared to the available experimental data. The full line
represents the experimental values.}\label{fig1}
\end{figure}

We apply our first principles approach to obtain E$_{2g}$
non-adiabatic phonon frequencies at ${\bf \Gamma}$ for a number of
GICs, for MgB$_2$ and for bulk $hcp$ Ti (see
footnote~\cite{CompDetails} for computational details). The
results are illustrated in Fig.~\ref{fig1} for GICs and are
collected in Tab.~\ref{table} for all the systems. In general huge
$\Delta\omega$ values ($
> 60 ~cm^{-1}$) are found in all layer compounds.
The most spectacular non-adiabatic effects are found in KC$_8$
($\Delta \omega=310 ~cm^{-1}$, 20\% of $\omega^A$) and in MgB$_2$
($\Delta \omega=230~cm^{-1}$, 30\% of $\omega^A$ ). Even in bulk
$Ti$ we find a significant shift, $\Delta \omega=12~cm^{-1}$,
(more than 8\% of $\omega^A$). From Fig.~\ref{fig1} it is evident
that experimental Raman
data~\cite{DresselReview,Guerard75,Eklund77,Eklund80,Eklund87} in
all GICs are in much closer agreement with $\omega^{NA}$ than with
$\omega^{A}$, whereas in MgB$_2$ Raman data lie in between the two
theoretical frequencies. In Ti, contrary to the layer metals, the
observed Raman frequencies~\cite{Titanium} are much closer to the
adiabatic value.

To further refine our theoretical model, we now compute the phonon
frequency in presence of a finite value of the electron-momentum
relaxation rate $\sigma$,
$\omega^\sigma=\omega^A+\Delta\omega^\sigma$.
$\Delta\omega^\sigma$ can be obtained using electron and hole
Green functions {\it dressed by the interaction}
(electron-electron, electron-phonon, and electron-defect) in the
calculation of the phonon self-energy and of the dynamical matrix
\cite{Zawadowski42,Maksimov95,Marsiglio92}. Assuming \emph{i)}
$\sigma$ independent of energy  and \emph{ii)} $n(\epsilon)$ and
$g(\epsilon_f)$ constants for bands within $\epsilon_F\pm\sigma$,
Refs.~\cite{Zawadowski42,Maksimov95,Marsiglio92} obtained:
\begin{equation}
\hbar\Delta\omega^\sigma \simeq \hbar\Delta\omega
\frac{(\hbar\omega^A)^2}{(\hbar\omega^A)^2+\sigma^2}. \label{eq6}
\end{equation}
Note that when $\hbar\omega^A\gg\sigma$,
$\Delta\omega^\sigma\simeq \Delta\omega$ and the experimentally
observed phonon frequency is $\omega^{NA}$, \emph{i.e.} the system
is in the purely NA regime. On the contrary, for
$\hbar\omega^A\ll\sigma$, $\Delta\omega^\sigma=0$ and the measured
phonon frequency is $\omega^A$, i.e. the system is completely
adiabatic.

The linewidth of an optical phonon mode at ${\bf \Gamma}$ is also
affected by the presence of a finite momentum-relaxation rate. The
decay of a phonon into non-interacting (undressed) electron-hole
pairs ($\sigma=0$) is forbidden for a zone-center optical mode if
the direct gap is larger than the phonon energy
($\hbar\omega\ll|\epsilon_{\textbf{k}e}-\epsilon_{\textbf{k}o}|$).
This condition being verified in all layer systems considered
here, thus a zero linewidth should be measured
\cite{Jepsen95,CalandraPRB05}. However in all stage-1 GICs and in
$MgB_2$, very large linewidths, hardly explainable in term of
anharmonicities, are measured. Analogously to the calculation of
$\omega^\sigma$, a finite momentum-scattering rate $\sigma$ can be
considered in the evaluation of the imaginary part of the phonon
self-energy
\cite{Zawadowski42,Maksimov95,Marsiglio92,Cappelluti06}. This
leads to the following expression for the phonon full linewidth at
half maximum due to the phonon decay in dressed electron-hole
pairs ($\gamma_{\sigma}^{EPC}$):
\begin{equation}
\frac{\gamma_\sigma^{EPC}}{2}
=\hbar\Delta\omega\frac{\sigma\hbar\omega^A}{(\hbar\omega^A)^2+\sigma^2}
\label{eq7}.
\end{equation}

\begin{figure}
\centerline{\includegraphics[width=8.25cm]{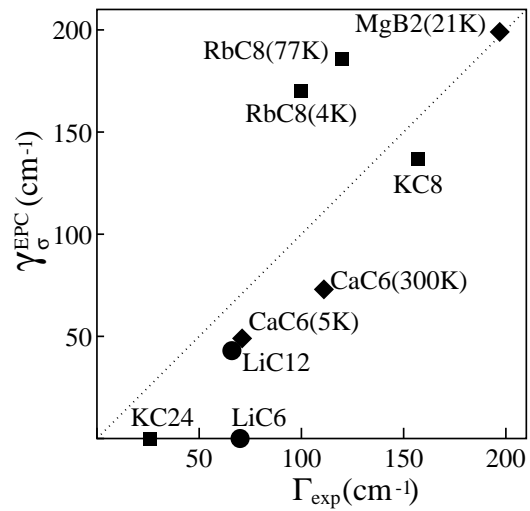}}
\caption{Calculated EPC linewidth (Eq.~\ref{eq7}) with respect to
the observed experimental ones.}\label{fig2}
\end{figure}

\begin{figure}
\centerline{\includegraphics[width=8.25cm]{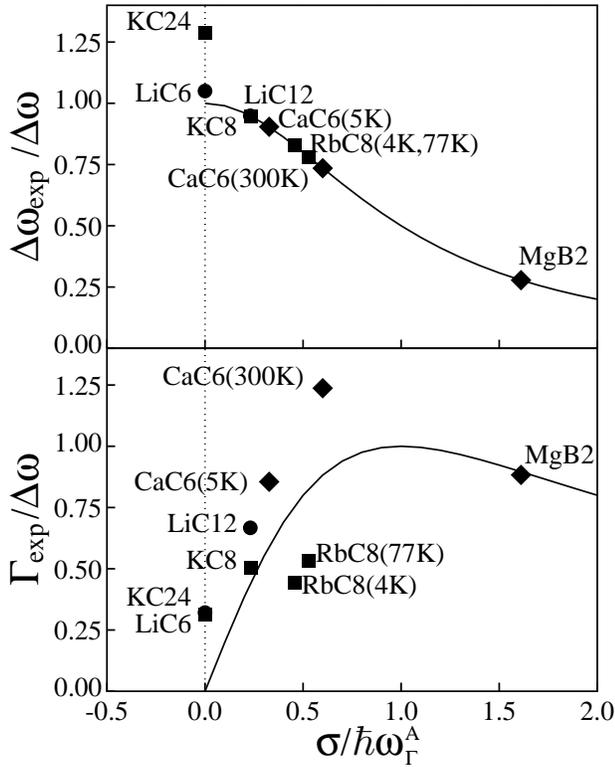}}
\caption{Upper panel:
$\hbar\Delta\omega_\sigma/\hbar\Delta\omega$ (Eq.~\ref{eq6}, solid
line), compared to measured frequencies
(symbols). Lower panel:
$\gamma_\sigma^{EPC}/\hbar\Delta\omega$ (Eq.~\ref{eq7}, solid
line), and measured linewidths (FWHM).}\label{fig3}
\end{figure}

Eq.~(\ref{eq6}) can be used to extract $\sigma$. Setting $\Delta
\omega^{\sigma}= \omega_{\rm exp}-\omega^A$, we obtain $\sigma$
from the inversion of Eq.~(\ref{eq6}). Then $\sigma$ is inserted
in Eq.~(\ref{eq7}) to determine $\gamma_{\sigma}^{EPC}$ and the
results are reported in Tab. \ref{table}. The comparison between
$\gamma_{\sigma}^{EPC}$ and the experimental Raman linewidths
$\Gamma_{\rm exp}$  is shown in Fig. \ref{fig2}. The agreement is
overall very good, except for LiC$_6$ and RbC$_8$. Note that the
experimental linewidth includes all sort of broadening (including
inhomogeneous effects), while $\gamma_{\sigma}^{EPC}$ includes
only the phonon decay into dressed electron-hole pairs. Thus when
$\Gamma_{\rm exp} \approx \gamma_{\sigma}^{EPC}$ it means that the
dominant broadening mechanism is the latter one. This is the case
in MgB$_2$ \cite{Cappelluti06,CalandraPhysC07}, and in most GICs.
Moreover, by comparing these results with the universal curves of
Eqs.~(\ref{eq6}) and~(\ref{eq7}) (Fig.~\ref{fig3}), we deduce that
our estimate of the momentum-relaxation time is a good indicator
of the degree of non-adiabaticity.

\begin{table}
\begin{tabular}{|c|c|c|c|c|c|c|c|}
\hline \hline & &  & & & & &\\
 &  $\omega^{A}$ & $\omega^{NA}$ & $\omega_{exp}$ &
$\Delta\omega$ & $\sigma$ & $\gamma^{EPC}_{\sigma}$ & $\Gamma_{exp}$ \\
\hline
$LiC_6$        &~1362~&~1580~&~1595~&~218~& 0     & 0 & 70\\
$LiC_{12}$     & 1492 & 1591 & 1590 & 99 &~151~& 20 & 66\\
\hline
$KC_8$        & 1223 & 1534 & 1522 & 311 & 245 &~120~& ~157~\\
$KC_{24}$     & 1488 & 1571 & 1599 & 83 & 0 & 0 & 26\\
\hline
$RbC_8$ {\scriptsize (4K)} & 1300 & 1525 & 1490 & 225 & 558 & 163 & 100\\
$RbC_8$ {\scriptsize (77K)}& 1300 & 1525 & 1480 & 225 & 650 & 180 & 120\\
\hline
$CaC_6$ {\scriptsize (5K)} & 1446 & 1529 & 1525 & 83 & 325 & 36 & 71\\
$CaC_6$ {\scriptsize (300K)}& 1446 & 1529 & 1511 & 83 & 761 & 68 & 111\\
\hline
$SrC_6$ & 1459 & 1530 & - & 71 & - & - & -\\
\hline
$BaC_6$ & 1462 & 1521 & - & 59 & - & - & -\\
\hline \hline
$MgB_2$ {\scriptsize (21K)} & 538 & 761 & 600 & 224 & 867 & 199 & 197\\
\hline \hline
$Mg$  & 122 & 123 & 122.5 & 1 & N/A & N/A & N/A\\
\hline
$Ti$  & 139 & 151 & 141 & 12 & N/A & N/A & N/A\\
\hline \hline
\end{tabular}
\caption{\label{table} Calculated adiabatic and non-adiabatic
E$_{2g}$ phonon frequencies, experimental frequencies, theoretical
frequency difference, calculated $\sigma$, calculated EPC
linewidth, experimental linewidth of several GICs and of $MgB_2$.
Experimental data are from
Refs.~\cite{Nancy07,Quilty02,DresselReview,Guerard75,Eklund77,Eklund80,Eklund87,Doll87}
Symbols indicate systems for which experimental data are
available, and are the same as in the figures. For $Mg$ and $Ti$
the determination of $\sigma$ from our theory is not applicable.
We note that since our best NA frequency of graphite is
1578$~cm^{-1}$, we have upshifted by 4$~cm^{-1}$ all the
adiabatic/NA frequencies of GICs.}
\end{table}

As an example, our results quantitatively indicate that $MgB_2$,
although being characterized by a huge $\Delta\omega$, has a
smaller $\Delta\omega^\sigma$ and a relatively short
momentum-relaxation time. It is thus a mostly adiabatic system,
even though $\Delta\omega^\sigma$ is still very large. On the
contrary, all GICs have relatively small $\sigma$'s, falling in
the left part of the universal curves of Fig.~\ref{fig3}, and are
thus mostly non-adiabatic. The result on $hcp$ titanium actually
indicates that all metals might have NA ${\bf \Gamma}$ frequencies
significantly different from the adiabatic ones, but that they
cannot be experimentally observed by Raman because the
condition~(\ref{eq1}) is not verified in experiments. In other
words, {\it non-adiabaticity is not a unique property of
low-dimensional or layer systems, and can occur even in perfectly
conventional metals}. Interestingly, despite the structural
similarity, in bulk $Mg$ no NA effect exists.

In conclusion, we provide a quantitative first-principles
theoretical framework to explain the difference between the
reported experimental E$_{2g}$ mode frequencies, and the
(ordinary) adiabatic calculated ones in several relevant layer
metals. We have shown that giant non-adiabatic effects occur in
these systems. Moreover, NA effects are in principle very relevant
even in bulk metals, although difficult to measure. Finally, we
have shown that the electron momentum-relaxation time can be
extracted from Raman peak positions and linewidths and is a good
indicator of the degree of non-adiabaticity of the system.

Calculations have been performed at the IDRIS French National
Computational Facility under the projet CP9-71387.


\bibliography{references}

\begin{thebibliography}{31}
\bibitem[{foo()}]{footnote1}
\bibinfo{note}{{For the sake of simplicity we consider atoms having identical
  masses M.}}
\end{thebibliography}

\end{document}